\documentclass[prb,aps,twocolumn,showpacs,floatfix,superscriptaddress]{revtex4}
\usepackage{graphicx}
\usepackage{amsmath,amsfonts}
\usepackage{amsbsy}

\newcommand{\slv}{\mbox{\sl v}}

\begin{document}

\title{Scattering of hole excitations in a one-dimensional spinless quantum liquid}

\author{K. A. Matveev}

\affiliation{Materials Science Division, Argonne National Laboratory,
  Argonne, Illinois 60439, USA}

\author{A. V. Andreev}

\affiliation{Department of Physics, University of Washington, Seattle,
  Washington 98195, USA}

\date{June 28, 2012}

\begin{abstract}

  Luttinger liquid theory accounts for the low energy boson
  excitations of one-dimensional quantum liquids, but disregards the
  high energy excitations.  The most important high energy excitations
  are holes which have infinite lifetime at zero temperature.  At
  finite temperatures they can be scattered by thermally excited
  bosons.  We describe the interaction of the hole with the bosons by
  treating it as a mobile impurity in a Luttinger liquid.  This
  approach enables us to evaluate the scattering probability at
  arbitrary interaction strength.  In general, the result is expressed
  in terms of the hole spectrum, its dependence on the density and
  momentum of the fluid, and the parameters of the Luttinger liquid
  Hamiltonian.  In the special case of Galilean invariant systems the
  scattering probability is expressed in terms of only the hole
  spectrum and its dependence on the fluid density.  We apply our
  results to the problem of equilibration of one-dimensional quantum
  liquids.

\end{abstract}


\maketitle

\section{Introduction}\label{sec:intro}

The Luttinger liquid theory describes low energy properties of one-dimensional
systems of interacting
fermions\cite{haldane1981luttinger,giamarchi2004quantum} or
bosons\cite{popov1972theory,haldane1981effective} in terms of excitations with
Bose statistics. The latter have an acoustic spectrum and correspond to long
wavelength density fluctuations. Despite the simplicity of the Hamiltonian,
the Luttinger liquid theory successfully describes the nontrivial power law
behavior\cite{dzyaloshinskii1974correlation} of various correlation functions
of one-dimensional systems at small frequencies and wavenumbers. Many of them,
such as the power law energy dependence of the tunneling density of states,
have been experimentally confirmed.\cite{bockrath1999luttinger}

\begin{figure}
\includegraphics[width=.45\textwidth]{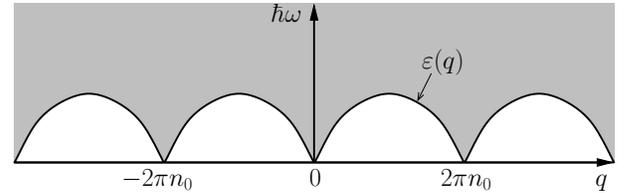}
\caption{ At a given momentum $\hbar q$ the energy of a one
   dimensional system is limited from below by the spectral edge
   $\varepsilon(q)$. Thus the imaginary part of the structure factor $S(q,
   \omega)$, describing the dissipation, vanishes outside the shaded region.
}
\label{fig:region}
\end{figure}

The power law behavior of correlators in one-dimensional systems is not
limited to small frequencies and wavenumbers.  For example, even at large wave
numbers $q\sim n_0$, where $n_0$ is the fluid density, the zero temperature
dynamic structure factor (the Fourier transform of the density-density
correlator) exhibits a power law singularity $S(q,\omega) \propto [\hbar
\omega-\varepsilon(q)]^\alpha\theta(\hbar \omega-\varepsilon(q))$. Its
location $\varepsilon(q)$ defines the spectral edge below which the system
cannot absorb excitations. It corresponds to the lowest energy state of the
system with momentum $\hbar q$.  The position of the spectral edge
$\varepsilon(q)$ is periodic in $q$ with the period $2\pi n_0$ (in the
spinless case considered here), as illustrated in Fig. \ref{fig:region}.

The nature of states corresponding to the spectral edge and periodicity of the
latter may be illustrated by the simple example of noninteracting fermions in
one dimension. Because the energy cost associated with transferring a particle
between the two Fermi points vanishes in the thermodynamic limit, the minimal
energy states with momenta differing by an integer multiple of $2\hbar k_F=
2\pi \hbar n_0$ are degenerate, resulting in the periodicity of the spectral
edge. It is thus sufficient to consider states with wave vectors in the
fundamental domain $0< q < 2\pi n_0$. The states at the spectral edge
correspond to hole excitations which are obtained by moving a fermion with the
wave vector $k_F-q$ to the right Fermi point $k=k_F$.

In the presence of interactions between the fermions the energy cost
associated with the transfer of a particle between the opposite Fermi
points still vanishes in the thermodynamic limit, and the spectral
edge $\varepsilon(q)$ remains periodic in $q$ with the period $2\pi
n_0$. A state at the spectral edge with a wave vector in the range
$0<q<2\pi n_0$ may again be viewed as the $q=0$ ground state with an
additional hole. However, the hole is now dressed by the interactions,
and its energy $\varepsilon(q)$ is renormalized. The above picture of
the spectral edge also applies to bosonic fluids, where holes are
known as Lieb's type-II excitations.\cite{lieb1963exact}

At $q\to 0$ and $q\to 2\pi n_0$ the energy $\varepsilon(q)$ is small
and the hole can be decomposed into the bosonic excitations of the
Luttinger liquid.  In contrast, the high energy holes with $q\sim n_0$
are not described by the Luttinger liquid theory and should be viewed
as distinct from the Luttinger liquid bosons.  As noted above, the
states with a single hole are lowest energy states of the system with
a given momentum.  Therefore they cannot decay into bosons and their
lifetime is infinite.

High energy hole excitations may be produced in the system by external
probes (e.g., optically or by tunneling) or by thermal
fluctuations. Therefore the problem of their dynamics is of
considerable interest.  Although the lifetime of the hole states is
infinite at zero temperature, the situation changes at $T>0$. In this
case the hole can scatter off thermally excited bosons.  Since the
resulting energy change $\delta\varepsilon\lesssim T$, a hole with
energy $\varepsilon(q)\gg T$ remains distinct from the Luttinger
liquid bosons and may be treated as a mobile impurity in a Luttinger
liquid.\cite{ogawa1992fermi, neto1996dynamics, pustilnik2006dynamic,
  khodas2007fermi, pereira2009spectral, imambekov_universal_2009,
  imambekov_phenomenology_2009, gangardt2010quantum,
  schecter_dynamics_2011}

\begin{figure}
\includegraphics[width=.4\textwidth]{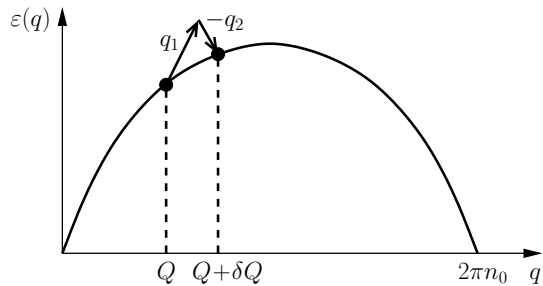}
\caption{Scattering of a hole at wavenumber
   $Q$ to a state $Q+\delta Q$ involves absorption of one bosonic excitations
   and emission of another.
}
\label{fig:scattering_process}
\end{figure}

In this paper we evaluate the probability of scattering of a hole by
the bosonic excitations.  The scattering event will be assumed to
change the wavenumber of the hole $Q$ by a small amount $\delta Q$.
In order to conserve both the energy and momentum of the system, the
hole has to absorb one boson and emit another, see
Fig.~\ref{fig:scattering_process}.  Such processes were first
considered by Castro Neto and Fisher\cite{neto1996dynamics} in the
context of the dynamics of mobile impurities in the Luttinger liquid.
The scattering probability near the top of the spectrum, $Q= \pi n_0$,
determines the rate of equilibration of one-dimensional quantum
liquids studied in Ref.~\onlinecite{matveev2011equilibration}.  In the
limit of strong repulsion the scattering processes of
Fig.~\ref{fig:scattering_process} were studied in the context of
equilibration of the one-dimensional Wigner
crystal,\cite{matveev2010equilibration, matveev2012rate} while for
weakly interacting bosons they are responsible for the decay of the
so-called dark soliton.\cite{gangardt2010quantum}  The mobile impurity
approach enables us to treat this problem at any interaction strength.
We express the scattering probability in terms of the hole spectrum
$\varepsilon(q)$ and its dependence on the density and velocity of the
fluid.  For Galilean invariant systems the result can be expressed in
terms of only the spectrum and its dependence of the fluid density.

The standard bosonization of Luttinger
liquids,\cite{haldane1981luttinger,giamarchi2004quantum} used in the
previous treatments of mobile impurities,\cite{ogawa1992fermi,
  neto1996dynamics, balents2000x, pustilnik2006dynamic,
  pereira2009spectral, khodas2007fermi, khodas2007dynamics,
  zvonarev2007spin, imambekov_universal_2009,
  imambekov_phenomenology_2009, zvonarev2009edge, kamenev2009dynamics,
  cheianov2008threshold} corresponds to the Eulerian description of
the liquid. In this approach the dynamical degrees of freedom
describing an element of the fluid are given as functions of time $t$
and the instantaneous position $x$ of the element in the laboratory
frame. Alternatively, the liquid may be described using Lagrangian
variables. In the latter approach the dynamical degrees of freedom are
labeled by the position $y$ of the fluid element in a reference state
of uniform density $n_0$.  For a liquid moving with a uniform velocity
the transformation from Eulerian to Lagrangian variables is equivalent
to a Galilean transformation from a laboratory frame to a reference
frame moving with the fluid.  Therefore one expects that Galilean
invariant systems are more naturally described in Lagrangian
variables.  This expectation is borne out: the use of Lagrangian
variables considerably simplifies evaluation of the scattering
amplitude.

On the other hand, Galilean invariance is not a universal property of
Luttinger liquids.  For instance, electrons in solids move in the
periodic potential of the lattice which gives rise to the band
structure of the spectrum.  In this case the quadratic spectrum
$p^2/2m$ required for Galilean invariance appears only near the band
edges. A similar situation arises when the concept of Luttinger liquid
is applied to spin chains and one-dimensional systems of cold atoms in
optical lattices.  In the absence of Galilean invariance Lagrangian
variables offer no obvious advantages.  To address this regime we
develop the theory of hole scattering using the standard Luttinger
liquid theory\cite{haldane1981luttinger, giamarchi2004quantum} based
on the Eulerian approach to fluid dynamics.

The paper is organized as follows. In Sec.~\ref{sec:Lagrangian} we
consider the Galilean invariant case, and develop theoretical
description of quantum liquids in Lagrangian variables.  We obtain the
scattering probability of the hole in terms of its spectrum.  In
Sec.~\ref{sec:Eulerian} we consider the general case, where Galilean
invariance is not assumed, and use the conventional approach based on
the Eulerian variables to obtain the scattering probability of the
excitation.  In Sec.~\ref{sec:discussion} we verify that the two
approaches give the same result in the case of a Galilean invariant
system and adapt our calculations to the problem of a massive mobile
impurity in a Luttinger liquid.  We also discuss the implications of
our results for the problem of equilibration of the Luttinger liquid.

\section{Scattering of holes in Galilean invariant systems}
\label{sec:Lagrangian}

In this section we consider systems which possess Galilean
invariance. Evaluation of the hole scattering probability is
simplified by developing the theory in Lagrangian
variables,\cite{landau1987fluid} which are frequently used to describe
one-dimensional flows in classical hydrodynamics.

\subsection{Hamiltonian of  a quantum liquid in Lagrangian variables}
\label{sec:bosonization_Lagrangian}

We first consider a liquid in which only low energy excitations are
present. These excitations are essentially sound waves with small wavenumbers
$q\ll n_0$. In this regime the discreteness of particles is not important and
the liquid may be described as a continuum in the spirit of hydrodynamics.

The Hamiltonian describing the long wavelength excitations may be
obtained by coarse-graining the fluid into small elements. The
Lagrangian coordinate $y$ of the fluid element corresponds to its
position in a reference state of uniform density $n_0$.  In the
presence of sound waves the particle density $n(y)$ inside each
element may deviate from $n_0$. The length of each element in the
reference state $\Delta y$ is assumed to be sufficiently small, $q
\Delta y \ll 1$, so that the variations of the fluid density inside it
are negligible.  On the other hand, the number of particles in the
element is assumed to be large, $n_0 \Delta y \gg 1$.

Taking advantage of the Galilean invariance we write the energy of the fluid
element as a sum of the kinetic energy of its center of mass motion and the
internal energy,
\begin{equation}\label{eq:energy_element}
    \Delta E=\frac{(\Delta P)^2}{2 m n_0 \Delta y} + U(n)\,  n_0 \Delta y.
\end{equation}
Here $\Delta P$ is the momentum of the element and $m$ is the mass of the
particles. In the absence of high energy excitations the internal energy is
given by the energy per particle in the ground state $U(n)$ multiplied by the
number of particles $n_0 \Delta y$.

The Hamiltonian of the liquid is written in terms of two dynamical variables,
the displacement of the fluid elements from their reference positions $u(y)$
and the conjugate momentum density $p(y)=\Delta P/\Delta y$, satisfying the
commutation relation $[u(y),p(y')]=i \hbar \delta (y-y')$. Summing the
energies (\ref{eq:energy_element}) of the fluid elements we obtain
\begin{equation}
  \label{eq:H_L}
  H_L=\int\left[
         \frac{p^2}{2mn_0} + n_0U(n)
        \right] dy.
\end{equation}
The Lagrangian variable $u(y)$ enters the Hamiltonian via the particle density
\begin{equation}\label{eq:n_u}
    n(y)=\frac{n_0}{1+u'(y)},
\end{equation}
where prime denotes the derivative. Equation (\ref{eq:n_u}) follows
immediately from the relation
\begin{equation}\label{eq:x_u}
    x=y+u(y)
\end{equation}
between the physical coordinate $x$ of the fluid element and the Lagrangian
variable $y$.

For the subsequent discussion it is sufficient to expand the Hamiltonian of
the liquid to third order in the deformation $u'$,
\begin{equation}
  \label{eq:H_0_approximate}
  H_L=\int\left(
         \frac{p^2}{2mn_0} + \frac{mn_0v^2}{2}\,u'^2 -\alpha u'^3
        \right) dy.
\end{equation}
The speed of sound in the liquid $v$ and the coefficient $\alpha$ in the
anharmonic term are given by
\begin{eqnarray}\label{eq:v}
    v(n_0)&=&\left\{ \frac{1 }{m} \left[2n_0U'(n_0)+n_0^2U''(n_0)\right]\right\}^{1/2}, \\
    \label{eq:alpha}
    \alpha (n_0) &=& n_0^2U'(n_0)+n_0^3U''(n_0)+\frac{1}{6}\, n_0^4U'''(n_0),
\end{eqnarray}
where derivatives are again denoted by prime.

The quadratic part of Eq.~(\ref{eq:H_0_approximate}) is the Luttinger liquid
Hamiltonian in Lagrangian variables.  It can be brought to the diagonal form
$\sum \hbar v|q|b_q^\dagger b_q$ by introducing the boson operators $b_q$ via
the standard procedure\cite{haldane1981luttinger,giamarchi2004quantum}
\begin{eqnarray}
  \label{eq:bosons_u}
  u(y)&=&\sum_q\sqrt{\frac{\hbar}{2mn_0Lv|q|}}\,
       \left(b_qe^{iqy}+b_q^\dagger e^{-iqy}\right),
\\
  \label{eq:bosons_p}
  p(y)&=&-i\sum_q\sqrt{\frac{\hbar mn_0v|q|}{2L}}\,
       \left(b_qe^{iqy}-b_q^\dagger e^{-iqy}\right).
\end{eqnarray}
where $L$ is the system size.

Because of the continuum approximation made in the description of the liquid
the sums in the above equations include only small wavevectors $|q|\ll q_0$,
where $q_0=1/\Delta y $. Therefore if a high energy hole is present in the
liquid, it needs to be treated as an additional entity.

\subsection{Description of the hole in Lagrangian variables}
\label{sec:hole_Lagrangian}

Let us now consider the fluid with a single high energy hole excitation. The
Lagrangian coordinate $Y$ of the hole refers to the fluid element containing
it.  As discussed in the introduction the hole excitation is obtained by
moving a particle from state $k_F-q$ to the Fermi point $k_F$. Thus the
presence of the hole has no effect on the mass of the fluid element, and the
expression for its kinetic energy given by the first term in
Eq.~(\ref{eq:energy_element}) remains unchanged. On the other hand, the
internal energy increases by the excitation energy $\varepsilon(q,n)$. (Here
we take into account the dependence of the latter on the fluid density $n$.)

It is important to note that in the presence of sound waves the physical size
of the fluid element $\Delta x$ differs from its size in the reference state,
$\Delta x=(n_0/n) \Delta y$. As a result the physical wavenumber $q$ differs
from the wavenumber $Q$ corresponding to the Lagrangian coordinate of the
excitation, $q=(n/n_0) Q$. Therefore it is convenient to introduce the
excitation spectrum $\epsilon (Q,n)$ with respect to the Lagrangian wavenumber
$Q$, which is related to the physical spectrum $\varepsilon (q,n)$ by
\begin{equation}\label{eq:spectrum_Lagrangian}
    \epsilon(Q, n)=\varepsilon(n Q/n_0,n).
\end{equation}

We now conclude that the presence of a hole at point $Y$ is accounted for by
the contribution to the Hamiltonian in the form
\begin{equation}\label{eq:H_hole}
    H_h=\epsilon(-i \partial_Y, n(Y)).
\end{equation}
The density $n(Y)$ at the location of the hole is affected by the
boson excitations, see Eqs.~(\ref{eq:n_u}) and (\ref{eq:bosons_u}).
This results in the interaction of the hole with the bosons.

\subsection{Scattering probability}
\label{sec:scattering_Lagrangian}

We now consider scattering processes shown in
Fig.~\ref{fig:scattering_process} and evaluate their rate $W_{Q,Q+\delta
  Q}$. To this end we expand the excitation energy to second order in $u'$
\begin{eqnarray}\label{eq:H_hole_expansion}
   \!\!\! \!\!\!\!\!\!\epsilon^{(2)}(Q,n(Y))&\!=\!&\epsilon_Q - n_0   \partial_n \epsilon_Q u'(Y) \nonumber \\
     &&+ \! \left[ n_0 \partial_n \epsilon_Q   + \frac{1}{2} n_0^2 \partial^2_n \epsilon_Q  \right]\!u'^2(Y).
\end{eqnarray}
In this equation and the subsequent results the hole energy
$\epsilon_Q=\epsilon(Q,n)$ and its partial derivatives are evaluated at
$n=n_0$. The corresponding Hamiltonian $H^{(2)}_h$ is obtained by substituting
$Q=- i \partial_Y$ and symmetrizing the operators. Because of the relation
(\ref{eq:bosons_u}) $H^{(2)}_h$ contains both linear and quadratic coupling
between the bosons and the hole.

The scattering process illustrated in Fig.~\ref{fig:scattering_process}
involves two bosons. Its amplitude may be obtained using perturbation theory
in the coupling between the hole and the bosons. One contribution to the
scattering amplitude arises from the quadratic coupling in the second line of
Eq.~(\ref{eq:H_hole_expansion}) taken in the first order of perturbation
theory. Keeping in mind Eq.~(\ref{eq:bosons_u}) one easily concludes that such
amplitude is proportional to the small momentum transfer $ \delta Q $.

Another contribution arises from the linear coupling term which scales as
$\sqrt{\delta Q}$. The process shown in Fig.~\ref{fig:scattering_process} is
realized in the second order perturbation theory in the linear coupling. Since
the energy denominator is proportional to $\delta Q$ one may naively expect
the respective amplitude to scale as $\left(\sqrt{\delta Q}\right)^2/\delta Q
= (\delta Q)^{0}$. However, there are two processes in which the two
participating bosons are created and destroyed in opposite order. Since the
corresponding energy denominators have opposite signs, the amplitudes of such
processes cancel each other in leading order in $\delta Q$. The remaining
subleading contribution again scales as $\delta Q$.

Finally there is a contribution in the second order perturbation theory, which
arises from the combination of the linear coupling of the bosons to the hole
and the $\alpha u'^3$ term in Eq.~(\ref{eq:H_0_approximate}). The latter
scales as $(\delta Q)^{3/2}$, resulting in the scattering amplitude linear in
$\delta Q$.

One can avoid the somewhat tedious calculation outlined above by performing a
unitary transformation $U^\dagger \big(H_L +H_h^{(2)} \big) U$ of the
Hamiltonian, which eliminates the linear coupling between the hole and the
bosons to leading order in $\delta Q$.  The operator $U$ should be chosen in
the form
\begin{equation}\label{eq:U}
    U=\exp\left( \frac{i}{\hbar} f_u u(Y) + \frac{i}{\hbar} f_p \int_{-\infty}^Y p(y) dy \right).
\end{equation}
Here the coefficients $f_u$ and $f_p$ are given by
\begin{equation}\label{eq:f}
    f_u= -\frac{n_0 v_Q \partial_n \epsilon_Q}{ v^2-v_Q^2}, \quad
    f_p=\frac{\partial_n \epsilon_Q}{m  (v^2 - v_Q^2)},
\end{equation}
where $v_Q$ is shorthand notation for the velocity of the hole,
\begin{equation}\label{eq:v_Q_physical}
    v_Q = v(Q,n)= \frac{1}{\hbar} \, \partial_Q \varepsilon (Q,n) .
\end{equation}

Because in the above unitary transformation $Q$ denotes the initial
momentum of the hole rather than the operator $-i \partial_Y$ the
linear coupling between the hole and the bosons is removed only to
leading order in $\delta Q$. However, in its absence the subleading
contribution, which scales as $(\delta Q)^{3/2}$, gives a negligible
correction to the scattering amplitude, $ [ ( \delta
Q)^{3/2}]^2/\delta Q = (\delta Q)^{2}$.  The dominant contribution to
the scattering amplitude arises from the terms in the transformed
Hamiltonian that couple the hole to second powers of boson fields,
$u'^2(Y)$ and $p^2(Y)$. This coupling has the form
\begin{eqnarray}
    && \bigg(-3\alpha f_p
                 +n_0\partial_n\epsilon_Q +\frac12 n_0^2\partial^2_n\epsilon_Q
                 -v_Q f_u-n_0\partial_n\slv_Q f_u 
\nonumber\\
           &&    - \frac{f_u^2}{2m_Q^*}\bigg)\, u'^2(Y)
                 - \frac{ f_p^2}{2 m^*_Q} \, p^2(Y),
\end{eqnarray}
where the momentum dependent effective mass $m^*_Q$ is defined in terms of the
curvature of the spectrum
\begin{equation}\label{eq:effective mass}
    \frac{1}{m^*_Q}=-\frac{1}{\hbar^2 }\, \partial_Q^2\epsilon_Q.
\end{equation}
In addition we introduced the notation
\begin{equation}\label{eq:v_Q}
    \slv_Q=\slv(Q,n)=v(nQ/n_0,n)
\end{equation}
for the velocity of the hole as a function of the Lagrangian wavenumber,
cf. Eq.~(\ref{eq:spectrum_Lagrangian}). Although $\slv_Q=v_Q$ at $n=n_0$,
their dependences on density are different. In particular,
\begin{equation}\label{eq:dn_slv}
    \partial_n \slv_Q =\partial_n v_Q - \frac{\hbar Q}{m_Q^* n_0}
\end{equation}
at $n=n_0$.

The matrix element for the scattering process in which a boson with momentum
$q_1$ is annihilated and a boson with momentum $q_2$ is created is obtained by
expressing the deformation $u'$ and the momentum density $p$ in the above
expression in terms of the boson creation and annihilation operators using
Eqs.~(\ref{eq:bosons_u}) and (\ref{eq:bosons_p}),
\begin{eqnarray}
   t_{q_1 q_2}&=& \frac{\hbar \sqrt{|q_1 q_2|}}{m n_0 L v} 
                \bigg[3\alpha f_p
                 -n_0\partial_n\epsilon_Q -\frac12 n_0^2\partial^2_n\epsilon_Q
                 +v_Q f_u
\nonumber \\
   &&  +n_0\partial_n\slv_Q f_u 
        +\frac{f_u^2-(m n_0 v f_p)^2}{2 m^*_Q}\bigg].
   \label{eq:t}
\end{eqnarray}
Here we assumed that $q_1 q_2<0$, see Fig.~\ref{fig:scattering_process}.

The coefficient $\alpha$ may be expressed in terms of the density dependence
of sound velocity $v$ with the aid of Eqs.~(\ref{eq:v}) and (\ref{eq:alpha}),
\begin{equation}
  \label{eq:alpha_final}
  3\alpha (n) = m n\left(v^2 +\frac12 n \partial_n v^2\right). 
\end{equation}
Using this relation and
Eq.~(\ref{eq:f}) we can express the scattering amplitude in terms of the
spectrum of the hole,
\begin{equation}
   t_{q_1 q_2}= \frac{\hbar\,  n_0 \sqrt{|q_1 q_2|}}{2m L v (v^2 - v_Q^2)}\, 
                \Upsilon_Q,
   \label{eq:t_Lagrange}
\end{equation}
where we introduced the notation
\begin{eqnarray}
    \Upsilon_Q &=& (\partial_n \epsilon_Q)\, 
     \partial_n (v^2-\slv_Q^2)- (v^2-v_Q^2) \, \partial_n^2\epsilon_Q 
\nonumber \\
     &&-  \frac{1}{m^*_Q}\, (\partial_n \epsilon_Q)^2 . 
\label{eq:Upsilon}
\end{eqnarray}

The scattering probability per unit time is given by the Fermi golden rule,
\begin{eqnarray}\label{eq:W}
     W_{Q, Q+ \delta Q} \!&\!=\!& \!\frac{ 2 \pi }{ \hbar}
    \! \sum_{q_1,q_2} \!|t_{q_1,q_2}|^2 N_{q_1} \!\left(N_{q_2}+1 \right) 
          \delta(q_1-q_2- \delta Q) 
\nonumber \\
    && \!\times \delta ( \epsilon_Q - \epsilon_{Q + \delta Q} +\hbar v |q_1| -\hbar v |q_2| ), \label{eq:golden_rule}
\end{eqnarray}
where $N_q$ is the occupation number of the boson state $q$.  We have
normalized the probability $W_{Q, Q+ \delta Q}$ in such a way that the total
scattering rate is given by the integral
\begin{equation}
  \label{eq:total_rate}
  W_Q=\int  W_{Q, Q+ \delta Q} \,d\delta Q.
\end{equation}
Using Eq.~(\ref{eq:t_Lagrange}) we immediately obtain
\begin{equation}\label{eq:W_Lagrange}
    W_{Q, Q+ \delta Q} = \frac{n_0^2  N_{q_1}(N_{q_2}+1 ) }{ 64\pi m^2 v^5 (v^2 -v_Q^2) }   \, \Upsilon^2_Q \,   (\delta Q)^2.
\end{equation}
Here $q_1$ and $q_2$ are given by the relations
\begin{equation}\label{eq:q_Q}
    q_1=\frac12\delta Q +\frac{v_Q}{2 v}|\delta Q|, 
\quad 
    q_2=-\frac12\delta Q +\frac{v_Q}{2 v}|\delta Q|, 
\end{equation}
arising from conservation of energy and momentum.

Equations (\ref{eq:Upsilon}) and (\ref{eq:W_Lagrange}) express the scattering
rate of a high energy hole excitation in a Galilean invariant quantum fluid in
terms of the hole spectrum (\ref{eq:spectrum_Lagrangian}). They are the main
result of this section.

\section{Scattering of holes in the absence of Galilean invariance}
\label{sec:Eulerian}

We now turn to the more general situation in which the quantum liquid is not
assumed to possess Galilean invariance.  As discussed above, in this case
Lagrangian variables offer no obvious advantages.  Therefore we apply the
standard (Eulerian) theory\cite{haldane1981luttinger, giamarchi2004quantum} of
the Luttinger liquid.

\subsection{Eulerian description of one-dimensional quantum liquids}
\label{sec:Euler_variables}

The standard theory of a Luttinger liquid describes the system by two bosonic
fields $\phi(x)$ and $\theta (x)$ satisfying the canonical commutation
relations
\begin{equation}\label{eq:commutation_phi_theta}
    [\phi(x), \nabla\theta(x')]=i \pi \delta(x-x').
\end{equation}
In contrast to the coordinate $y$ in the Lagrangian approach of
Sec.~\ref{sec:Lagrangian}, the coordinate $x$ denotes the real space position,
see Eq.~(\ref{eq:x_u}).

The field $\phi$ is defined in terms of the particle density
\begin{equation}\label{eq:phi_density}
    n(x)=n_0 +\frac{1}{\pi} \nabla \phi(x).
\end{equation}
The field $\theta$ accounts for the motion of the liquid. The latter may be
characterized by the momentum per particle
\begin{equation}\label{eq:kappa}
    \kappa(x)=\frac{1}{2}\left( \frac{1}{n(x)}\, p(x) + p(x)\, \frac{1}{n(x)}\right).
\end{equation}
Here the momentum density
\begin{equation}\label{eq:momentum_density}
    p(x)= \frac{1}{2}\sum_l \left[ p_l  \delta(x-x_l) + \delta(x-x_l) \, p_l\right]
\end{equation}
is defined in terms of the coordinates $x_l$ and momenta $p_l$ of the physical
particles.  Using the commutators $[x_l, p_{l'}]=i \hbar \, \delta_{ll'}$ and
definition of the particle density
\begin{equation}\label{eq:density_definition}
    n(x)=\sum_l \delta (x-x_l),
\end{equation}
one easily obtains the commutation relation
\begin{equation}\label{eq:commutator_density_kappa}
    [n(x),\kappa(x')]= - i \hbar \nabla \delta(x-x').
\end{equation}
Comparing this relation with Eqs.~(\ref{eq:commutation_phi_theta}) and
(\ref{eq:phi_density}) we identify
\begin{equation}
  \label{eq:theta_kappa}
   \kappa (x)= - \hbar  \nabla \theta (x).
\end{equation}
Thus the gradient of the boson field $\theta$ determines momentum per particle
in the fluid.  

It is worth mentioning that in the Galilean invariant case
$\kappa (x) = m V(x)$ and $\nabla \theta (x)$ gives the expression
\begin{equation}
    V(x)= -\frac{\hbar}{m} \nabla \theta (x)
\end{equation}
for the fluid velocity.  The relation between the fields $\phi(x)$,
$\theta(x)$ and the Lagrangian variables $u(y)$, $p(y)$ used in
Sec.~\ref{sec:Lagrangian} is given by Eqs.~(\ref{eq:n_u}),
(\ref{eq:phi_density}), (\ref{eq:kappa}), and (\ref{eq:theta_kappa}),
as well as Eq.~(\ref{eq:x_u}) which expresses the physical coordinate
$x$ in terms of the Lagrangian coordinate $y$.

\subsection{Hamiltonian of the liquid in the presence of the hole}
\label{sec:Hamiltonian_Euler}

The standard form of the Hamiltonian of a Luttinger liquid is
\begin{equation}\label{eq:H_0}
    H_0=\frac{\hbar v}{2\pi} \int dx \left[ K (\nabla \theta)^2
      +K^{-1}(\nabla \phi)^2\right]
\end{equation}
(see, e.g., Sec. 3.1 of Ref.~\onlinecite{giamarchi2004quantum}).  Here
the velocity $v$ and the dimensionless Luttinger liquid parameter $K$
depend on the density of the particles and the interactions between
them.  This Hamiltonian describes noninteracting bosons with an
acoustic spectrum.  It can be brought to the form
\begin{equation}\label{eq:H_0_bosons}
    H_0=\sum_q \hbar v |q| \left( b^\dagger_q b_q +\frac{1}{2}\right)
\end{equation}
with the help of the relations
%\begin{subequations}
\begin{eqnarray}
  \label{eq:bosons_phi}
 \!\!\!\!\!\!\! \nabla \phi (x) \!&\!=\!&-i\sum_q\sqrt{\frac{\pi K|q|}{2L}}\, \mathrm{sgn} (q)
       \!\left(b_qe^{iqx}-b_q^\dagger e^{-iqx}\right)\!,
\\
  \!\!\!\!\!\!\!\nabla \theta(x)\!&\!=\!& i \sum_q\sqrt{\frac{\pi |q| }{2 K L}}\,
       \left(b_q e^{iqx} - b_q^\dagger e^{-iqx}\right),  \label{eq:bosons_theta}
\end{eqnarray}
which express the fields $\phi$ and $\theta$ in terms of the boson creation and
annihilation operators $b^\dagger_q$ and $b_q$.

The Hamiltonian $H_0$ describes the low energy properties of the
system and may be viewed as the fixed point Hamiltonian in the
renormalization group sense. The leading irrelevant perturbation is
given by cubic in the boson fields
corrections,\cite{haldane1981luttinger, footnote}
\begin{equation}
    H_\alpha=\int d x \left[\alpha_\theta ( \nabla \phi )(\nabla \theta)^2 + \alpha_\phi (\nabla \phi)^3 \right].\label{eq:H_alpha_Euler}
\end{equation}
Here we assumed that the system is invariant under inversion, $x\to
-x$. In this case the Hamiltonian cannot contain odd powers of $\nabla
\theta$. The parameters $\alpha_\theta$ and $\alpha_\phi$ can be
expressed in terms of density dependent parameters of the fixed point
Hamiltonian (\ref{eq:H_0}).  Indeed, according to
Eq.~(\ref{eq:phi_density}) a small change of density $\delta n$ shifts
$\nabla \phi \to \nabla \phi + \pi \delta n$. As a result the cubic
perturbation (\ref{eq:H_alpha_Euler}) generates a small correction to
the quadratic Hamiltonian (\ref{eq:H_0}). The corresponding change in
its parameters is $\delta (\hbar v K/2\pi) =\pi \alpha_\theta\delta
n$, and $\delta (\hbar v /2 \pi K) = 3\alpha_\phi \delta n$. We thus
find the expressions
\begin{equation}
\label{eq:alpha_relations}
  \alpha_\theta = \frac{\hbar}{ 2\pi^2}\,  \partial_n\left( v K\right), \quad
  \alpha_\phi = \frac{\hbar}{6\pi^2}\, \partial_n \left( \frac{v}{K}\right)
\end{equation}
for the parameters of the Hamiltonian (\ref{eq:H_alpha_Euler}).

Let us now consider a hole excitation at point $X$. In general its energy $\varepsilon$ depends not only on the wave number $Q=-i \partial_X$ but also on the density of the liquid $n(X)$ and its motion quantified by $\kappa (X)$. If the spectrum of the hole in a uniform liquid $\varepsilon(Q, n, \kappa)$ is known, the correction to the Hamiltonian of the system due to the presence of the hole may be written as
\begin{equation}\label{eq:H_h_Euler}
    H_h=\varepsilon(-i \partial_X, n (X), \kappa (X)).
\end{equation}
The right hand side here is assumed to be symmetrized with respect to the operators $-i \partial_X$, $n(X)$, and $\kappa(X)$. In view of the relations (\ref{eq:phi_density}) and (\ref{eq:theta_kappa}) this Hamiltonian describes the hole interacting with the boson fields $\phi$ and $\theta$.

\subsection{Scattering probability}
\label{sec:scattering_Euler}

The Hamiltonian of the liquid in the presence of the hole given by Eqs.~(\ref{eq:H_0}), (\ref{eq:H_alpha_Euler}) and (\ref{eq:H_h_Euler}) is similar to that obtained in Lagrangian variables for the Galilean invariant case, Eqs. (\ref{eq:H_0_approximate}) and (\ref{eq:H_hole}). The main difference is that the spectrum of the hole in the Eulerian description depends not only on the fluid density $n$ but also on its momentum $\kappa$. The rate $W_{Q,Q+\delta Q}$ of scattering of the hole by bosons, Fig.~\ref{fig:scattering_process}, can be found by repeating the steps of Sec.~\ref{sec:scattering_Lagrangian}. First we perform a unitary transformation $U^\dagger H U$ of the Hamiltonian with the operator
\begin{equation}\label{eq:U_Euler}
    U=\exp\left[ i f_\phi \phi(X) + i f_\theta \theta (X) \right].
\end{equation}
Upon such a transformation the Hamiltonian takes the form $H_0+ H_\alpha + \tilde{H}_h$ where
\begin{eqnarray}
    \tilde{H}_h&=&\varepsilon(-i \partial_X + f_\phi \nabla \phi + f_\theta \nabla \theta  , n_0+ \nabla \phi/\pi , - \hbar \nabla \theta ) \nonumber \\
    &&+ \hbar v K f_\phi \nabla \theta + \frac{\hbar v}{ K } f_\theta \nabla \phi \nonumber \\
    && + \pi f_\theta \left[ \alpha_\theta (\nabla \theta )^2 + 3 \alpha_\phi (\nabla \phi)^2\right] \nonumber \\
    && + 2\pi f_\phi \alpha_\theta \nabla \phi \nabla \theta .\label{eq:H_h_tilde}
\end{eqnarray}
Here the gradients of the boson fields are evaluated at point $X$.

The parameters $f_\phi$ and $f_\theta$ should be chosen such that the linear terms in the expansion of Eq.~(\ref{eq:H_h_tilde}) in powers of $\nabla \phi$ and $\nabla \theta$ vanish to leading order in
$-i \partial_X -Q$, where $Q$ is the initial wavenumber of the hole, Fig.~\ref{fig:scattering_process}. This yields
\begin{eqnarray}
  f_\phi &=& \frac{1}{\pi \hbar K}\frac{v_Q K \partial_n \varepsilon_Q +  \pi \hbar   v \, \partial_\kappa \varepsilon_Q}{v^2 - v_Q^2},  \label{eq:f_phi}\\
  f_\theta &=& - \frac{1}{\pi \hbar}\frac{v K \partial_n \varepsilon_Q + \pi \hbar   v_Q \partial_\kappa \varepsilon_Q}{v^2 - v_Q^2} ,\label{eq:f_theta}
\end{eqnarray}
where we used the shorthand notations $\varepsilon_Q=\varepsilon(Q,n,\kappa)$ and $v_Q= \partial_Q \varepsilon_Q/\hbar$. In Eqs.~(\ref{eq:f_phi}),  (\ref{eq:f_theta}), and the subsequent results $\varepsilon_Q$ and its partial derivatives are evaluated at $n=n_0$ and $\kappa=0$.

With the above choice of $f_\phi$ and $f_\theta$ the Hamiltonian (\ref{eq:H_h_tilde}) takes the form
\begin{eqnarray}
  \tilde{H}_h &=& (\nabla \theta)^2 \!\left( \pi \alpha_\theta f_\theta -
            \frac{\hbar^2 f_\theta^2}{2 m_Q^*}  +\frac{\hbar^2}{2} \partial_\kappa^2 \varepsilon_Q - \hbar^2 f_\theta \partial_\kappa v_Q\right) \nonumber  \\
        &&+(\nabla \phi)^2\!\left( 3\pi \alpha_\phi f_\theta -\frac{\hbar^2 f_\phi^2}{2 m_Q^*}  +\frac{\partial_n^2 \varepsilon_Q }{2\pi^2} + \frac{\hbar f_\phi}{\pi} \partial_n v_Q\right) \nonumber  \\
         &&+  \, \varepsilon_Q + \ldots . \label{eq:tilde H_h_expansion}
\end{eqnarray}
Here we omitted linear in $\nabla \phi$ and $\nabla \theta$ terms with
coefficients small in $-i \partial_X-Q$, the quadratic term proportional to
$(\nabla \phi) (\nabla \theta)$, and higher powers of $\nabla \phi$ and
$\nabla \theta$.  Such perturbations do not affect the amplitude of the
scattering process depicted in Fig.~\ref{fig:scattering_process} to leading
order in $\delta Q$.  In Eq.~(\ref{eq:tilde H_h_expansion}) the effective mass
of the hole $m_Q^*$ is defined by
\begin{equation}\label{eq:m_Q_Euler}
    \frac{1}{m^*_Q}=- \frac{1}{\hbar^2} \, \partial_Q^2 \varepsilon (Q, n_0, 0).
\end{equation}
We now substitute relations (\ref{eq:bosons_phi}) and (\ref{eq:bosons_theta}) into Eq.~(\ref{eq:tilde H_h_expansion}) and extract the matrix element $t_{q_1q_2}$ corresponding to the scattering process in Fig.~\ref{fig:scattering_process}. Using Eqs.~(\ref{eq:alpha_relations}), (\ref{eq:f_phi}), and (\ref{eq:f_theta}) we express the matrix element $t_{q_1q_2}$ in the form
\begin{equation}\label{eq:t_Upsilon_Euler}
    t_{q_1 q_2}=\frac{K}{2\pi L} \frac{\sqrt{|q_1 q_2|}}{v^2-v_Q^2} \, \Upsilon_Q,
\end{equation}
where
\begin{eqnarray}
  \Upsilon_Q &=& -\frac{1}{m_Q^*}  \left[ (\partial_n \varepsilon_Q)^2 -\left(\frac{\pi \hbar}{K}\right)^2 (\partial_\kappa \varepsilon_Q )^2\right] \nonumber \\
  && -(v^2 - v_Q^2) \left[ \partial_n^2 \varepsilon_Q - \left(\frac{\pi \hbar}{K}\right)^2 \partial_\kappa^2 \varepsilon_Q\right] \nonumber \\
  && + 2 \left( \frac{\pi \hbar}{K} \partial_\kappa v_Q -\frac{v}{K} \partial_n K\right)\!\!
  \left( v \partial_n \varepsilon_Q + v_Q \frac{\pi \hbar}{K} \partial_\kappa \varepsilon_Q\right)
  \nonumber \\
  && - 2 (\partial_n v_Q) \left[ v_Q \partial_n \varepsilon_Q + v \frac{\pi \hbar}{K} \partial_\kappa \varepsilon_Q \right]
  . \label{eq:Upsilon_Euler}
\end{eqnarray}
Substituting Eq.~(\ref{eq:t_Upsilon_Euler}) into the Fermi golden rule
expression (\ref{eq:golden_rule}) we obtain the scattering rate
\begin{equation}\label{eq:W_Euler}
    W_{Q, Q+ \delta Q} = \frac{K^2  N_{q_1}(N_{q_2}+1 ) }{64 \pi^3 \hbar^2  v^3 (v^2 -v_Q^2) }   \, \Upsilon^2_Q \,   (\delta Q)^2,
\end{equation}
where $q_1$ and $q_2$ are again given by Eq.~(\ref{eq:q_Q}).

Equations (\ref{eq:W_Euler}) and (\ref{eq:Upsilon_Euler}) express the scattering rate of the hole in terms of its spectrum $\varepsilon (Q, n, \kappa)$ and the parameters of the Luttinger liquid. They are the main result of this section.

\section{Discussion of the results}
\label{sec:discussion}

In sections \ref{sec:Lagrangian} and \ref{sec:Eulerian} we obtained
expressions for the rate $W_{Q,Q+\delta Q}$, of scattering processes
illustrated in Fig.~\ref{fig:scattering_process}.  In
Sec.~\ref{sec:Lagrangian} we considered Galilean invariant systems and
obtained the hole scattering rate in the form of Eqs.~(\ref{eq:W_Lagrange})
and (\ref{eq:Upsilon}). Our consideration in Sec.~\ref{sec:Eulerian} did not
assume Galilean invariance and yielded the scattering rate in a somewhat more
complicated form, Eqs.~(\ref{eq:W_Euler}) and (\ref{eq:Upsilon_Euler}).  Below
we compare these results and discuss their applications to the problem of
equilibration of one-dimensional quantum liquids and to the problem of
dynamics of a mobile impurity in a Luttinger liquid.

\subsection{Galilean invariant systems}
\label{sec:Galilean_invariant_case}

To obtain the expressions (\ref{eq:W_Lagrange}) and (\ref{eq:Upsilon}) for the
scattering rate we described the liquid in Lagrangian coordinates. The state
of the hole was parameterized by the Lagrangian wave number conjugate to the
Lagrangian coordinate of the hole $Y$, rather than the conventional wave
number conjugate to the spatial coordinate $X$. Accordingly, the excitation
energy $\epsilon_Q=\epsilon(Q,n)$ and velocity $\slv_Q=\slv (Q,n)$ were
expressed in terms of the Lagrangian wavenumber using
Eqs.~(\ref{eq:spectrum_Lagrangian}) and (\ref{eq:v_Q}).

To illustrate this point let us consider the simple special case of
noninteracting fermions. Recalling that the hole excitation is created by
moving a fermion from state $k_F-Q$ to state $k_F$ we find its energy and
velocity in the form
\begin{equation}\label{eq:varepsilon_free_fermions}
    \varepsilon_Q=\frac{\hbar^2}{2 m}\, Q(2\pi n -Q), 
    \quad v_Q=\frac{\hbar}{m}(\pi n -Q).
\end{equation}
To express the energy and velocity in terms of the Lagrangian wave number we
substitute $Q \to n Q/n_0$, see Eqs.~(\ref{eq:spectrum_Lagrangian}) and
(\ref{eq:v_Q}), and find
\begin{equation}\label{eq:epsilon_free_fermions}
     \epsilon_Q=\frac{\hbar^2 n^2}{2 m n_0^2}\, Q(2\pi n_0 -Q),
\quad \slv_Q=\frac{\hbar n}{m n_0}(\pi n_0 -Q).
\end{equation}
Substituting (\ref{eq:epsilon_free_fermions}) into (\ref{eq:Upsilon}) we
obtain $\Upsilon_Q=0$. This is the expected result, as no scattering of
excitations may occur in a system of noninteracting fermions. An erroneous
substitution of the $\varepsilon_Q$ and $v_Q$ given by
Eq.~(\ref{eq:varepsilon_free_fermions}) for $\epsilon_Q$ and $\slv_Q$ into
Eq.~(\ref{eq:Upsilon}) would result in $\Upsilon_Q \neq 0$.

Although less common than the Eulerian variables in hydrodynamics,
Lagrangian variables arise naturally in the theory of elasticity. In
particular, in the case of a one-dimensional Wigner crystal
(i.e. quantum anharmonic chain) the Lagrangian coordinate is
essentially the number of a site of the Wigner lattice.  Thus the
phonon spectrum $\omega_Q$ of the Wigner crystal is naturally
expressed in terms of the Lagrangian wavenumber. The weakness of
interactions between the phonons in the Wigner crystal enables one to
develop a microscopic theory of scattering of a high energy phonon by
acoustic modes.\cite{matveev2010equilibration, matveev2012rate} On the
other hand, the Wigner crystal is simply the limiting case of a
quantum fluid in the regime of extremely strong repulsion between the
particles.  Thus the scattering of a high energy phonon can also be
studied phenomenologically using the approach of
Sec.~\ref{sec:Lagrangian}.  Indeed, substituting the phonon energy
$\epsilon(Q,n)=\hbar\omega_Q$ into Eq.~(\ref{eq:Upsilon}) we recover
the result of Ref.~\onlinecite{matveev2012rate}.

Experimentally, it is easier to measure the excitation spectrum
$\varepsilon(Q,n)$ as a function of the conventional wavenumber. To obtain the
scattering rate in terms of $\varepsilon(Q,n)$ using Eq.~(\ref{eq:W_Lagrange})
one should substitute the relations (\ref{eq:spectrum_Lagrangian}) and
(\ref{eq:v_Q}) into Eq.~(\ref{eq:Upsilon}). This yields,
\begin{eqnarray}
    \Upsilon_Q &=& (\partial_n \varepsilon_Q)\, \partial_n (v^2-v_Q^2)- (v^2-v_Q^2) \, \partial_n^2\varepsilon_Q \nonumber \\
     &&- \, \frac{1}{m^*_Q}\, (\partial_n \varepsilon_Q)^2 + \frac{\hbar^2}{m^*_Q} \frac{v^2 Q^2}{n_0^2} \nonumber \\
     &&+  \frac{2 \hbar v Q}{n_0} \left( v_Q \partial_n v - v \partial_n v_Q\right) . \label{eq:Upsilon_varepsilon}
\end{eqnarray}
As before, the right hand side here is evaluated at $n=n_0$.  Substitution of
the expressions (\ref{eq:varepsilon_free_fermions}) for the energy and the
velocity of the hole in a noninteracting Fermi gas into
Eq.~(\ref{eq:Upsilon_varepsilon}) gives the correct result $\Upsilon_Q=0$.

Our discussion of hole scattering in Sec.~\ref{sec:Eulerian} did not assume
Galilean invariance of the system. The resulting scattering rate
$W_{Q,Q+\delta Q}$ is given by Eqs.~(\ref{eq:W_Euler}) and
(\ref{eq:Upsilon_Euler}). For systems that possess Galilean invariance these
results should agree with those of Sec.~\ref{sec:Lagrangian},
Eqs.~(\ref{eq:W_Lagrange}) and (\ref{eq:Upsilon}).  The Luttinger liquid
parameter $K$ in Galilean invariant systems can be expressed in terms of the
particle density $n$ and the velocity $v$ of excitations as
\begin{equation}\label{eq:K_Galilean}
    K=\frac{\pi \hbar n}{m v}.
\end{equation}
Substituting this expression into Eq.~(\ref{eq:W_Euler}) and setting $n=n_0$
we recover Eq.~(\ref{eq:W_Lagrange}). To demonstrate that the resulting
scattering rates are equal, we also need to derive the expression
(\ref{eq:Upsilon}) for $\Upsilon_Q$ from Eq.~(\ref{eq:Upsilon_Euler}).

In Sec.~\ref{sec:Eulerian} the hole was described by the dependence of its
energy on the wavenumber $Q$, particle density $n$ and the momentum per
particle $\kappa$. In Galilean invariant systems $\kappa=m V$, where $V$ is
the fluid velocity. The energy of the excitation $\varepsilon(Q, n,\kappa)$ in
the moving fluid differs from its energy $\varepsilon(Q,n)$ in the stationary
fluid by $\hbar Q V$ (see e.g. Ref.~\onlinecite{landauStat_Mech_II}),
resulting in
\begin{equation}\label{eq:Galilean_transformation}
    \varepsilon(Q,n, \kappa) =\varepsilon(Q, n) + \hbar Q \frac{\kappa }{m}.
\end{equation}
Substituting Eqs.~(\ref{eq:K_Galilean}) and (\ref{eq:Galilean_transformation})
into Eq.~(\ref{eq:Upsilon_Euler}) we obtain the result
(\ref{eq:Upsilon_varepsilon}), which is equivalent to Eq.~(\ref{eq:Upsilon}).

\subsection{Hole vs.\ particle-hole excitation}

Throughout this paper we considered hole excitations created by moving a
particle from a state $k_F-Q$ to the Fermi point $k_F$.  On the other hand,
our discussion of hole scattering in Sec.~\ref{sec:Eulerian} did not rely on
this physical picture and thus should be applicable to a hole excitation
obtained by removing a particle from the system.  
The wavenumber $\tilde Q$ of such an excitation is related to $Q$ by
\begin{equation}
  \label{eq:tilded_Q}
  \tilde Q(n,\kappa)=Q-k_F^R(n,\kappa),
\end{equation}
where $k_F^R$ is the wavenumber of the particle at the right Fermi point.
Similarly, its energy $\tilde\varepsilon$ is related to $\varepsilon$ by
\begin{equation}
  \label{eq:tilded_varepsilon}
  \varepsilon(Q,n,\kappa)=\mu^R(n,\kappa) 
                         +\tilde\varepsilon(Q-k_F^R(n,\kappa),n,\kappa),
\end{equation}
where $\mu^R$ is the energy of the particle at the right Fermi point.  Since
the two physical pictures of the hole excitation are equivalent, one should
expect to find the same scattering rate for the particle-hole excitation with
energy $\varepsilon(Q,n,\kappa)$ as for the hole (missing particle) with
energy $\tilde\varepsilon(Q-\pi n_0,n,\kappa)$.

To verify such a feature of Eq.~(\ref{eq:Upsilon_Euler}) we need to obtain the
expressions for $k_F^R(n,\kappa)$ and the derivatives of $\mu^R(n,\kappa)$.
To this end we express $n$ and $\kappa$ in terms of the numbers $N^R$ and
$N^L$ of the right- and left-moving particles in a uniform system,
\begin{equation}
  \label{eq:n_and_kappa}
  n=\frac{N^R+N^L}{L},
\quad
  \kappa=\pi\hbar\,\frac{N^R-N^L}{L}.
\end{equation}
Then from $k_F=(2\pi/L)N^R$ we find
\begin{equation}
  \label{eq:k_F^R}
  k_F^R(n,\kappa)=\pi n + \frac\kappa\hbar.
\end{equation}
To find the derivatives of $\mu^R(n,\kappa)$ we substitute
$\nabla\phi=\pi(n-n_0)$ and $\nabla\theta=-\kappa/\hbar$ into the Hamiltonian
of the liquid given by Eqs.~(\ref{eq:H_0}) and (\ref{eq:H_alpha_Euler}).
Differentiating the resulting expression for the energy of uniform liquid with
respect to $N^R$ we find the expression
\begin{eqnarray}
  \label{eq:mu^R}
  \mu^R(n,\kappa)&=&\mu^R(n_0,0)+\frac{\pi\hbar v}{K}(n-n_0)+vK\kappa
\nonumber\\
                 &&+\frac{\pi\alpha_\theta}{\hbar^2}
                    [\kappa^2+2\pi\hbar(n-n_0)\kappa]
\nonumber\\
                 &&+3\pi^3\alpha_\phi(n-n_0)^2+\ldots
\end{eqnarray}
valid to second order in $n-n_0$ and $\kappa$.  From the first line one then
immediately obtains the first derivatives of $\mu^R$ in the form
\begin{equation}
  \label{eq:mu^R_first_derivatives}
  \partial_n\mu^R=\pi\hbar \,\frac{v}{K},
\quad
  \partial_\kappa\mu^R=vK.
\end{equation}
The most convenient expression for the second derivative $\partial_n^2\mu^R$
is obtained simply by differentiating the first of the expressions
(\ref{eq:mu^R_first_derivatives}).  To find $\partial_\kappa^2\mu^R$ we notice
that according to the second line of Eq.~(\ref{eq:mu^R}) we have
$\partial_\kappa^2\mu^R=(\pi\hbar)^{-1}\partial_n(\partial_\kappa \mu^R)$.
Then from Eq.~(\ref{eq:mu^R_first_derivatives}) we obtain
\begin{equation}
  \label{eq:mu^R_second_derivatives}
  \partial_n^2\mu^R=\pi\hbar\, \partial_n\frac{v}{K},
\quad
  \partial_\kappa^2\mu^R=\frac{\partial_n (vK)}{\pi\hbar}.
\end{equation}
Using the expressions (\ref{eq:k_F^R}), (\ref{eq:mu^R_first_derivatives}), and
(\ref{eq:mu^R_second_derivatives}) one can show that $\Upsilon_Q$ obtained by
the substitution of $\epsilon(Q,n,\kappa)$ in the form
(\ref{eq:tilded_varepsilon}) into Eq.~(\ref{eq:Upsilon_Euler}) coincides with
that obtained by simply replacing
$\varepsilon(Q,n,\kappa)\to\tilde\varepsilon(Q-\pi n_0,n,\kappa)$ in
Eq.~(\ref{eq:Upsilon_Euler}).

\subsection{Scattering of a mobile impurity in a Luttinger liquid}
\label{sec:impurity}

Apart from a hole excitation created by either moving a particle to the Fermi
point or removing it from the system, the results of Sec.~\ref{sec:Eulerian}
apply to any mobile impurity in a Luttinger liquid.  Our result for the
scattering rate given by Eqs.~(\ref{eq:W_Euler}) and (\ref{eq:Upsilon_Euler})
is applicable regardless of Galilean invariance of the system, but requires
the knowledge of the energy of the mobile impurity $\varepsilon(Q,n,\kappa)$
as a function of the density and momentum of the liquid.  Similar to the case
of a hole excitation, Eq.~(\ref{eq:Galilean_transformation}), this relation
simplifies in the presence of Galilean invariance.  One should note, however,
that unlike an intrinsic particle-hole excitation, a foreign particle has a
non-zero mass $M$, which affects the dynamics of the liquid.  The dependence
of the energy $\varepsilon_p(V)$ of the impurity with momentum $p$ on the
velocity $V$ of the fluid can be obtained by comparing the expressions for the
energy of the system in the stationary frame and that moving with the fluid,
\begin{equation}
  \label{eq:Baym_Ebner_relation}
  \varepsilon_p(V)=\varepsilon_{p-MV}(0)+pV-\frac12 MV^2,
\end{equation}
see Ref.~\onlinecite{PhysRev.164.235}.  

Denoting the momentum of the impurity $p=\hbar Q$, the velocity of the fluid
$V=\kappa/m$, and expanding to second order in $\kappa$, we obtain
\begin{eqnarray}
  \label{eq:Galilean_transformation_with_M}
  \varepsilon(Q,n,\kappa) &\simeq&
 \varepsilon(Q,n) +(\hbar Q - Mv_Q)\frac{\kappa}{m}
\nonumber\\
         && -\frac{M}{2}\left(1+\frac{M}{m_Q^*}\right)\frac{\kappa^2}{m^2}. 
\end{eqnarray}
Substituting this expression into Eq.~(\ref{eq:Upsilon_Euler}) we find
\begin{eqnarray}
  \label{eq:Upsilon_Galilean_with_M}
    \Upsilon_Q &=& (\partial_n \varepsilon_Q)\, \partial_n (v^2-v_Q^2)-
    (v^2-v_Q^2)\bigg(\partial_n^2\varepsilon_Q +\frac{Mv^2}{n_0^2}\bigg)
\nonumber \\
     && -\frac{1}{m^*_Q}\left(\partial_n
       \varepsilon_Q-\frac{Mv^2}{n_0}\right)^2 
      +\frac{\hbar^2Q^2}{m^*_Q} \frac{v^2}{n_0^2} 
\nonumber \\
     && +\frac{2v}{n_0}  (\hbar Q-Mv_Q)
        \left( v_Q \partial_n v - v \partial_n v_Q\right).  
\end{eqnarray}
This expression can be viewed as a generalization of our earlier result
(\ref{eq:Upsilon_varepsilon}) to the case of impurity with non-zero mass $M$.
The full expression for the rate of scattering of such an impurity is given by
the combination of Eqs.~(\ref{eq:Upsilon_Galilean_with_M}) and
(\ref{eq:W_Lagrange}).  The same result can, of course, be obtained by
generalizing the approach based on Lagrangian variables,
Sec.~\ref{sec:Lagrangian}, to the case of a massive particle.  The respective
calculation is outlined in the Appendix.

As a simple check we consider an impurity completely decoupled from the
liquid.  In this case the scattering probability must vanish, and we expect to
obtain $\Upsilon_Q=0$.  This is easily verified by substituting
$\partial_n\varepsilon_Q=0$, $v_Q=\hbar Q/M$ and $m_Q^*=M$ into
Eq.~(\ref{eq:Upsilon_Galilean_with_M}).

The problem of dynamics of a mobile impurity in a Galilean invariant
fluid was recently addressed by Schecter, Gangardt, and
Kamenev.\cite{schecter_dynamics_2011} Comparing our expression
(\ref{eq:Upsilon_Galilean_with_M}) with the Eq.~(75) of
Ref.~\onlinecite{schecter_dynamics_2011} we find an agreement,
provided that their matrix element $\Gamma_{+-}$ is related to
$\Upsilon_Q$ as
\begin{equation}
  \label{eq:Gamma_vs_Upsilon}
  \Gamma_{+-}=\frac{n_0}{mv}\frac{\Upsilon_Q}{(v^2-v_Q^2)^2}.
\end{equation}
The authors of
Ref.~\onlinecite{schecter_dynamics_2011} characterized the impurity by the
number of particles $N$ it expels from the fluid and the superfluid phase
$\Phi$.  The relation (\ref{eq:Gamma_vs_Upsilon}) was obtained by expressing
$N$ and $\Phi$ in terms of the energy and velocity of the impurity with the
help of Eq.~(21) of Ref.~\onlinecite{schecter_dynamics_2011}.

\subsection{Dissipative dynamics of holes and equilibration of
  one-dimensional quantum liquids}
\label{sec:dynamics}

In Secs.~\ref{sec:Lagrangian} and \ref{sec:Eulerian} we have evaluated the
rate $W_{Q,Q+\delta Q}$ of scattering of a hole excitation by the low energy
bosons.  Our results (\ref{eq:W_Lagrange}) and (\ref{eq:W_Euler}) enable one
to study the dynamics of the hole excitation, provided the occupation numbers
$N_q$ of bosonic states are known.  In the most interesting case when the
Luttinger liquid is in equilibrium at temperature $T$, the rate $W_{Q,Q+\delta
  Q}$ falls off exponentially at $|\delta Q|\gg T/(v-|v_Q|)$.  If a single
hole with energy $\varepsilon\gg Tv/(v-|v_Q|)$ is excited in the Luttinger
liquid, the collisions with bosons give rise to a gradual change of its
momentum at the rate
\begin{equation}
  \label{eq:force_definition}
  F=\left\langle \frac{d}{dt}\hbar Q\right\rangle=
    \int\hbar\delta Q\, W_{Q,Q+\delta Q}\, d\delta Q.
\end{equation}
Using our most general result (\ref{eq:W_Euler}) for the scattering rate we
obtain the force acting on the hole in the form
\begin{equation}
  \label{eq:force_result}
  F=-\frac{2\pi K^2\Upsilon_Q^2T^4}{15\hbar^5}
      \frac{(v^2+v_Q^2)v_Q}{(v^2-v_Q^2)^5v^2}.
\end{equation}
The negative sign in this expression indicates that as a result of scattering
by the bosons the wavenumber of the hole approaches one of the Fermi points,
$Q=0$ or $Q=2\pi n_0$, i.e., the hole is eventually absorbed into the boson
gas. 

As we stated in Sec.~\ref{sec:impurity} our results can also be applied to a
mobile impurity in a Luttinger liquid.  An expression for the force acting on
such an impurity in the case of Galilean invariant system was found in
Ref.~\onlinecite{schecter_dynamics_2011}.  Using Eqs.~(\ref{eq:K_Galilean})
and (\ref{eq:Gamma_vs_Upsilon}) we find that our result
(\ref{eq:force_result}) recovers Eq.~(73) of
Ref.~\onlinecite{schecter_dynamics_2011}.

The collisions of the hole with the bosons result in a stochastic motion which
should be described in terms of the hole distribution function $f_Q$.
Assuming again that the bosons are in thermal equilibrium and $Q$ is
sufficiently far from the Fermi points, the evolution of the distribution
function is controlled by the collision integral
\begin{equation}
  \label{eq:collision_integral}
  \frac{d f_Q}{dt}=-\int dQ' [f_QW_{Q,Q'}-f_{Q'} W_{Q',Q}].
\end{equation}
One can now use our results for $W_{Q,Q+\delta Q}$ to study the evolution of
the distribution function towards the equilibrium
$f_Q^{(0)}=e^{-\varepsilon(Q)/T}$.

The collision integral (\ref{eq:collision_integral}) takes a particularly
simple form for $Q$ in the vicinity of $Q_0=\pi n_0$, where the energy
$\varepsilon(Q)$ takes the maximum value, Fig.~\ref{fig:scattering_process}.
In this case the typical change of energy in a collision
$\delta\varepsilon\sim v_Q\delta Q\sim (v_Q/v)T\ll T$.  Thus the distribution
function changes very little after each collision.  This enables one to bring
the collision integral (\ref{eq:collision_integral}) to the Fokker-Planck form
\begin{equation}
  \label{eq:Fokker-Planck}
  \frac{d}{dt} f=-\partial_Q\left(A(Q)f-\frac12\partial_Q[B(Q)f]\right),
\end{equation}
where 
\begin{eqnarray}
  \label{eq:A}
  A(Q)&=&\int\delta Q\, W_{Q,Q+\delta Q}\,d\delta Q,
\\
  \label{eq:B}
  B(Q)&=&\int(\delta Q)^2\, W_{Q,Q+\delta Q}\,d\delta Q,
\end{eqnarray}
see, e.g., Ref.~\onlinecite{van_Kampen}.

The Fokker-Planck equation for the evolution of the hole distribution
function has been applied earlier to the problem of equilibration of a
one-dimensional system of interacting
electrons.\cite{micklitz2010transport, matveev2011equilibration,
  matveev2010equilibration, matveev2012rate} At low temperatures the
collisions lead to relatively fast thermalization of excitations near
each Fermi point.  However, the full equilibration of the system
includes exchange of particles between the right- and left-moving
branches, which equilibrates the respective chemical potentials.  This
process involves diffusion of a hole excitation in momentum space from
one Fermi point to the other.\cite{matveev2011equilibration} In the
case of arbitrary interaction strength, when the one-dimensional
system is treated as a Luttinger liquid, the equilibration rate is
expressed as\cite{matveev2011equilibration}
\begin{equation}
  \label{eq:equilibration_rate}
  \tau^{-1}= \frac{3\hbar n_0^2  B}{\sqrt{2\pi m^*T}} 
            \left(\frac{\hbar v}{T}\right)^3  e^{-\Delta/T}.
\end{equation}
Here $\Delta$, $m^*$, and $B$ are given, respectively, by the hole energy
$\varepsilon(Q_0)$, effective mass $m_{Q_0}^*$, and $B(Q_0)$ at the maximum
$Q_0=\pi n_0$.  Our results for $W_{Q,Q+\delta Q}$ allow us to use
Eq.~(\ref{eq:B}) to evaluate $B$ for any Luttinger liquid, regardless of the
presence of Galilean invariance.  Using Eq.~(\ref{eq:W_Euler}) we find
\begin{equation}
  \label{eq:B_result}
  B=\frac{4\pi}{15}\,\frac{K^2\Upsilon_{Q_0}^2}{\hbar^7v^{10}}\, T^5.
\end{equation}
In the case of a Galilean invariant system, Eq.~(\ref{eq:B_result}) recovers
the result of Ref.~\onlinecite{matveev2011equilibration}.

\subsection{Integrable models}
\label{sec:integrable}

The excitation spectrum of a one-dimensional quantum liquid can be obtained
exactly for integrable models.  A simple example of such a model is that of
spinless fermions of mass $m$ interacting with potential decaying as inverse
square of the distance between particles, $V(r)=g/r^2$.  The hole spectrum of
this Calogero-Sutherland model is given by\cite{sutherland2004beautiful}
\begin{equation}
  \label{eq:varepsilon_CS}
  \varepsilon(Q,n) = \frac{\hbar^2\lambda}{2m}\,Q(2\pi n-Q),
\end{equation}
where $\lambda(\lambda-1)=gm/\hbar^2$.  The velocity of holes is easily
obtained from Eq.~(\ref{eq:varepsilon_CS}),
\begin{equation}
  \label{eq:velocity_CS}
  v(Q,n) = \frac{\hbar\lambda}{m}(\pi n -Q),
\end{equation}
and the velocity of bosonic excitations $v=v(0,n)=\pi\hbar\lambda n/m$.  Upon
substitution of these values into the expression (\ref{eq:Upsilon_varepsilon})
one finds $\Upsilon_Q=0$, which points to the absence of scattering of holes by
bosonic excitations.

The absence of scattering of excitations is widely believed to be a universal
property of integrable models.\cite{mazets2008breakdown, gangardt2010quantum,
  schecter_dynamics_2011, tan2010relaxation, imambekov2011review,
  pereira2009spectral, matveev2010equilibration, matveev2011equilibration,
  matveev2012rate} Indeed, integrability implies the presence of a large
number of integrals of motion, which precludes scattering of excitations.
Given the expressions (\ref{eq:W_Lagrange}) and (\ref{eq:W_Euler}) for the
scattering probability, we expect that $\Upsilon_Q=0$ for any integrable
model.

\section{Summary}
\label{sec:summary}

To summarize, in this paper we developed a phenomenological theory of
scattering of hole excitations in one-dimensional spinless quantum liquids.
We expressed the scattering rate in terms of the spectrum of the hole and its
dependence on the fluid density and velocity. We considered liquids which may
or may not possess Galilean invariance.

Our approach is based on the concept of Luttinger liquid and thus applies at
any strength of the interactions between the particles.  We used two
alternative descriptions of the liquid, based on either Lagrangian or Eulerian
variables.  The former has the advantage of simplicity when applied to
Galilean invariant systems.  The latter is somewhat more complicated but
provides a more natural description of systems without Galilean invariance.

Our most general result for the scattering probability of a hole $W_{Q,Q+\delta
  Q}$ is given by Eqs.~(\ref{eq:W_Euler}) and (\ref{eq:Upsilon_Euler}).  It
simplifies considerably for Galilean invariant systems.  In this case
$W_{Q,Q+\delta Q}$ is given by Eq.~(\ref{eq:W_Lagrange}) where $\Upsilon_Q$
may be expressed in two equivalent ways.  It takes the form (\ref{eq:Upsilon})
if the energy of the hole is given as a function of the Lagrangian wavenumber.
This description arises naturally in the limit of strong repulsion between
particles, when the system forms a Wigner
crystal.\cite{matveev2010equilibration, matveev2012rate} Alternatively, if the
energy of the hole is known as a function of the Eulerian wavenumber the
expression (\ref{eq:Upsilon_varepsilon}) for $\Upsilon_Q$ is more convenient.

In our theory the hole excitation is treated as a mobile impurity.
Consequently, our result for the scattering probability (\ref{eq:W_Euler}) and
(\ref{eq:Upsilon_Euler}) applies not only to the hole, but any impurity,
including a foreign particle introduced into the system.  In the Galilean
invariant case, dynamics of such a particle was studied recently by Schecter,
Gangardt, and Kamenev.\cite{schecter_dynamics_2011}  We have verified that
their expression for the force acting on the particle agrees with our
Eq.~(\ref{eq:force_result}).

The scattering of intrinsic hole excitations controls equilibration of
one-dimensional quantum liquids.  Apart from fundamental importance, the
equilibration determines the conductance of long uniform quantum
wires.\cite{micklitz2010transport, matveev2011equilibration1} Previous
calculations of the equilibration rate\cite{ matveev2011equilibration,
  matveev2010equilibration, matveev2012rate} assumed Galilean invariance.  Our
discussion in Sec.~\ref{sec:dynamics} extends these results to the general
case.

\begin{acknowledgments}
  The authors are grateful to A.~Kamenev for discussions.  This work was
  supported by the U.S. Department of Energy under Contracts
  No. DE-AC02-06CH11357 and DE-FG02-07ER46452.

\end{acknowledgments}

\appendix*

\section{Description of a mobile impurity in Lagrangian variables}

In Sec.~\ref{sec:impurity} we discussed scattering of a massive mobile
impurity in a Luttinger liquid.  Our approach was based on the conventional
description of the system in terms of Eulerian variables.  In this Appendix we
derive the main result (\ref{eq:Upsilon_Galilean_with_M}) using Lagrangian
variables, see Sec.~\ref{sec:Lagrangian}.  We start by generalizing the
expression (\ref{eq:H_hole}) for the Hamiltonian of the excitation to account
for non-vanishing mass $M$ of the mobile impurity.  

Following the approach of Sec.~\ref{sec:Lagrangian}, we consider an element of
the liquid of length $\Delta y$ which contains the impurity.  We write the
energy of the element as a sum of the kinetic energy of center of mass of the
system and the energy in the center of mass frame,
\begin{equation}
  \label{eq:Delta_H_with_M}
  \Delta H = \frac{(\Delta P)^2}{2(mn_0\Delta y+M)} 
             + U(n)n_0 \Delta y + \epsilon(P_Y/\hbar,n),
\end{equation}
cf.~Eqs.~(\ref{eq:energy_element}) and (\ref{eq:H_hole}).  Expanding
Eq.~(\ref{eq:Delta_H_with_M}) to first order in $M/(mn_0\Delta y)\ll 1$, we
obtain
\begin{eqnarray}
  \label{eq:Delta_H_with_M_expanded}
  \Delta H &=& \frac{(\Delta P)^2}{2mn_0\Delta y} 
             + U(n)n_0 \Delta y 
\nonumber\\
           && +\epsilon(P_Y/\hbar,n)
              -\frac12 M \left(\frac{\Delta P}{mn_0\Delta y}\right)^2.
\end{eqnarray}
Here $P_Y$ and $\Delta P$ are the momenta conjugated to the Lagrangian
coordinate $Y$ of the impurity and the center of mass coordinate $R$ of the
element, respectively.  Because of the non-vanishing mass of the impurity, the
latter is no longer equivalent to the displacement $u$ of the fluid,
\begin{equation}
  \label{eq:center_of_mass}
  R=\frac{mn_0\Delta y [y+u(y)]+M[Y+u(Y)]}{mn_0\Delta y+M}.
\end{equation}
Here $y$ is the Lagrangian coordinate of the fluid element containing the
impurity, $y-\Delta y/2<Y<y+\Delta y/2$.  A small change in the positions of
the element $du$ and the particle $dY$ results in the shift of the center of
mass by
\begin{equation}
  \label{eq:dR}
  dR=du+\frac{M[1+u'(Y)]}{mn_0\Delta y+M}dY.
\end{equation}
Unlike the case of $M=0$ considered in Sec.~\ref{sec:Lagrangian}, the center
of mass position is affected by the motion of the impurity.  Despite that, the
momentum $\Delta P$ is still expressed in terms of the operator $\partial_u$,
%Therefore the momenta can be expressed as
\begin{equation}
  \label{eq:Delta_P}
  \Delta P = -i\hbar\partial_R\big|_{Y=\rm const} = -i\hbar\partial_u =
  \Delta y\, p(y),
\end{equation}
where $p(y)$ is the momentum of the liquid per unit length
introduced in Sec.~\ref{sec:bosonization_Lagrangian}.  On the other hand, the
momentum $P_Y=-i\hbar\partial_Y|_{R=\rm const}$ of the relative motion of the
impurity and the liquid differs from the momentum $\hbar
Q=-i\hbar\partial_Y|_{u=\rm const}$ of the impurity in a stationary liquid,
\begin{equation}
  \label{eq:P_Y}
  P_Y = \hbar Q
         -\frac{M[1+u'(Y)]}{mn_0\Delta y+M}(-i\hbar\partial_u).
\end{equation}
We now neglect in the denominator the mass of the impurity $M$
compared to the much larger mass $mn_0\Delta y$ of the fluid element and use
Eq.~(\ref{eq:Delta_P}) to exclude $\partial_u$.  This yields
\begin{equation}
  \label{eq:P_Y_final}
  P_Y = \hbar Q - \frac{M}{m}\,\frac{p(Y)}{n(Y)},
\end{equation}
see Eq.~(\ref{eq:n_u}).  Since $p(Y)$ is the momentum density of the liquid,
at $u'=0$ one can interpret the last term in the right-hand side as $-MV$,
where $V$ is the velocity of the liquid.

Our theory is constructed in terms of the displacement of the fluid $u$ rather
than the center of mass coordinate $R$.  Thus we substitute
Eqs.~(\ref{eq:Delta_P}) and (\ref{eq:P_Y_final}) into the expression
(\ref{eq:Delta_H_with_M_expanded}) for the Hamiltonian of the fluid element.
Then the first line of Eq.~(\ref{eq:Delta_H_with_M_expanded}) gives the energy
of the liquid without the impurity, cf.~Eq.~(\ref{eq:H_L}), whereas the second
line gives the Hamiltonian of the impurity in the form
\begin{equation}
  \label{eq:H_i}
  H_i = \epsilon\left(Q-\frac{M}{m}\, \frac{p(Y)}{\hbar n(Y)}, n(Y)\right) 
       -\frac12 M \left[\frac{p(Y)}{mn_0}\right]^2.
\end{equation}
At $M=0$ this Hamiltonian recovers that of the massless hole,
Eq.~(\ref{eq:H_hole}).

To obtain the probability $W_{Q,Q+\delta Q}$ of scattering of the impurity
from state $Q$ to $Q+\delta Q$ per unit time, we repeat the steps outlined in
Sec.~\ref{sec:scattering_Lagrangian} using the Hamiltonian (\ref{eq:H_i})
instead of (\ref{eq:H_hole}).  The first step is to perform the unitary
transformation $U^\dagger(H_L+H_i)U$ of the form (\ref{eq:U}).  The latter
transforms the relevant operators as follows
\begin{eqnarray}
  \label{eq:transformed_operators}
  U^\dagger u'(y) U &=& u'(y) + f_p\delta(y-Y),
\\
  U^\dagger p(y) U &=& p(y) + f_u\delta(y-Y),
\\
  U^\dagger Q U &=& Q +\frac{1}{\hbar}f_u u'(Y)+\frac{1}{\hbar}f_p p(Y).
\end{eqnarray}
We then choose the coefficients $f_u$ and $f_p$ such that the linear in the
bosonic fields $u$ and $p$ contribution
\begin{eqnarray}
  \label{eq:linear_contribution}
  H^{(1)}&=&(mn_0v^2 f_p + v_Q f_u -n_0 \partial_n \epsilon_Q)\, u'(Y)
\nonumber\\
  &&+\left(\frac{f_u}{mn_0}+v_Q f_p-\frac{Mv_Q}{mn_0}\right) p(Y)
\end{eqnarray}
to the transformed Hamiltonian $U^\dagger(H_L+H_i)U$ vanishes.  This gives
\begin{equation}
  \label{eq:f_u_and_f_p}
  f_u=-\frac{v_Q(n_0\partial_n\epsilon_Q-Mv^2)}{v^2-v_Q^2},
\quad
  f_p=\frac{n_0\partial_n\epsilon_Q-Mv_Q^2}{mn_0(v^2-v_Q^2)}.
\end{equation}

The next step is to expand the transformed Hamiltonian to second order in
bosonic fields,
\begin{eqnarray}
  \label{eq:quadratic_contribution}
\hspace{-2em}
  H^{(2)} &=& \Bigg[-3\alpha f_p
                 +n_0\partial_n\epsilon_Q +\frac12 n_0^2\partial^2_n\epsilon_Q
\nonumber\\
           &&    -v_Q f_u-n_0\partial_n\slv_Q f_u 
                 -\frac{f_u^2}{2m_Q^*}
             \Bigg]u'^2(Y)
\nonumber\\
           && -\Bigg[
               \frac{(mn_0 f_p-M)^2}{2m_Q^*}+\frac{M}{2}
               \Bigg] \frac{p^2(Y)}{m^2n_0^2}+\ldots.
\end{eqnarray}
Here ellipses denote the omitted terms of the form $u'(Y)p(Y)$ which generate
coupling of the impurity to two bosons on the same branch and are therefore
not relevant for our scattering problem, see
Fig.~\ref{fig:scattering_process}.

We now substitute the expressions (\ref{eq:bosons_u}) and (\ref{eq:bosons_p})
into (\ref{eq:quadratic_contribution}) and extract the matrix element
corresponding to annihilation of boson $q_1$ and creation of boson $q_2$,
\begin{eqnarray}
  \label{eq:tq1q2}
   t_{q_1 q_2}&=& \frac{\hbar \sqrt{|q_1 q_2|}}{m n_0 L v} 
                 \bigg(3\alpha f_p
                 -n_0\partial_n\epsilon_Q -\frac12 n_0^2\partial^2_n\epsilon_Q
\nonumber \\
   &&  +v_Q f_u+n_0\partial_n\slv_Q f_u 
                 +\frac{f_u^2}{2m_Q^*}
\nonumber \\
   &&  -\frac{(mn_0 f_p-M)^2v^2}{2m_Q^*}-\frac{Mv^2}{2}
 \bigg).
\end{eqnarray}
Substituting relations (\ref{eq:alpha_final}) and (\ref{eq:f_u_and_f_p}) we
bring this expression to the form (\ref{eq:t_Lagrange}) with
\begin{eqnarray}
  \label{eq:Upsilon_Lagrangian}
  \Upsilon_Q&=&(\partial_n \epsilon_Q)\,\partial_n (v^2-\slv_Q^2) 
     - (v^2-v_Q^2) \left(\partial_n^2\epsilon_Q +\frac{Mv^2}{n_0^2}\right)
\nonumber \\
     &&  -\frac{1}{m^*_Q} 
          \left(\partial_n \epsilon_Q-\frac{Mv^2}{n_0}\right)^2 
\nonumber\\
     &&  +\frac{2M}{n_0}vv_Q(v\partial_n \slv_Q-v_Q\partial_n v). 
\end{eqnarray}
This expression generalizes our earlier result (\ref{eq:Upsilon}) to the case
of a massive mobile impurity.  One can now substitute
Eqs.~(\ref{eq:spectrum_Lagrangian}) and (\ref{eq:v_Q_physical}) to express
$\Upsilon_Q$ in terms of the physical energy $\varepsilon$ of the impurity in
the liquid.  This procedure transforms the expression
(\ref{eq:Upsilon_Lagrangian}) to the form (\ref{eq:Upsilon_Galilean_with_M}).


\begin{thebibliography}{35}
\bibitem{footnote} In some lattice systems, such as that of
  one-dimensional fermions at half filling, perturbations in the form
  of cosines of bosonic fields appear.  In this case the problem of
  scattering of hole excitations is more complicated.  We leave this
  problem for future studies.
\end{thebibliography}
\end{document}